\documentclass[aps,prc,nokeys,nopacs,twocolumn,preprintnumbers,amsmath,amssymb,nofootinbib,floats
]{revtex4-1}
\usepackage{feynmp}
\usepackage{graphicx}
\usepackage{amssymb}
\usepackage{amsmath}
\usepackage{bm}
\usepackage{multirow}
\usepackage{color}
\usepackage{colordvi}
\usepackage{verbatim}
\usepackage{bbm}
\usepackage{bigstrut}
\usepackage{fix-cm}
\usepackage[labelfont={normalsize},subrefformat=parens,caption=false]{subfig}
\usepackage{enumerate}

\newcommand{\ad}[1]{a_{#1}}
\newcommand{\ac}[1]{a^{\dagger}_{#1}}

\newcommand{\bra}[1]{\langle #1 \vert}
\newcommand{\ket}[1]{\vert #1 \rangle}

\newcommand{\nuc}[2]{\ensuremath{{}^{#2}\mathrm{#1}}}
\newcommand{\keV}{\ensuremath{\,\text{keV}}}
\newcommand{\MeV}{\ensuremath{\,\text{MeV}}}

\begin{document}

\title{Non-observable nature of the nuclear shell structure.\\
Meaning, illustrations and consequences}

\author{T.~Duguet$^{1,2,3}$}
\email{thomas.duguet@cea.fr}

\author{H.~Hergert$^{3,4}$}
\email{hergert@nscl.msu.edu}

\author{J.~D.~Holt$^{5,6,7}$}
\email{jholt@triumf.ca}

\author{V.~Som\`a$^{1}$}
\email{vittorio.soma@cea.fr}

\affiliation{$^{1}$Centre de Saclay, IRFU/Service de Physique Nucl\'eaire,
F-91191 Gif-sur-Yvette, France}

\affiliation{$^{2}$Department of Physics and Astronomy,
Michigan State University,
East Lansing, Michigan 48824-1321, USA}

\affiliation{$^{3}$National Superconducting Cyclotron Laboratory,
Michigan State University,
East Lansing, Michigan 48824-1321, USA}

\affiliation{$^{4}$The Ohio State University, Columbus, Ohio 43210, USA}

\affiliation{$^{5}$TRIUMF, 4004 Wesbrook Mall, Vancouver, BC, V6T 2A3, Canada}
\affiliation{$^6$Institut f\"ur Kernphysik, Technische Universit\"at
Darmstadt, 64289 Darmstadt, Germany}
\affiliation{$^7$ExtreMe Matter Institute EMMI, GSI Helmholtzzentrum f\"ur
Schwerionenforschung GmbH, 64291 Darmstadt, Germany}
\begin{abstract}
\begin{description}
\item[Background]
The concept of single-nucleon shells constitutes a basic pillar of our understanding of nuclear structure. Effective single-particle energies (ESPEs) introduced by French and Baranger represent the most appropriate tool to relate many-body observables to a single-nucleon shell structure.  As briefly discussed in [T. Duguet, G. Hagen, Phys. Rev. C {\bf 85}, 034330 (2012)], the dependence of ESPEs on one-nucleon transfer probability matrices makes them purely theoretical quantities that ``run'' with the non-observable resolution scale $\lambda$ employed in the calculation.  
\item[Purpose]
Given that ESPEs provide a way to interpret the many-body problem in terms of simpler theoretical ingredients, the goal is to specify the terms, i.e. the exact sense and conditions, in which this interpretation can be conducted meaningfully. 
\item[Methods]
While the nuclear shell structure is both scale and scheme dependent, the present study focuses on the former. A detailed discussion is provided to illustrate the scale (in)dependence of observables and non-observables and the reasons why ESPEs, i.e. the shell structure, belong to the latter category.
State-of-the-art multi-reference in-medium similarity renormalization group and self-consistent Gorkov Green's function many-body calculations are employed to corroborate the formal analysis. This is done by comparing the behavior of several observables and of non-observable ESPEs (and spectroscopic factors) under (quasi) unitary similarity renormalization group transformations of the Hamiltonian parameterized by the resolution scale $\lambda$.
\item[Results]
The formal proofs are confirmed by the results of ab initio many-body calculations in their current stage of implementation. In practice, the unitarity of the similarity transformations is broken due to the omission of induced many-body interactions beyond three-body operators and to the non-exact treatment of the many-body Schr\"odinger equation. The impact of this breaking is first characterized by quantifying the artificial running of observables over a (necessarily) finite interval of $\lambda$ values. Then, the genuine running of ESPEs is characterized and shown to be convincingly larger than the one of observables (which would be zero in an exact calculation). 
\item[Conclusions]
The non-observable nature of the nuclear shell structure, i.e. the fact that it constitutes an intrinsically theoretical object with no counterpart in the empirical world, must be recognized and assimilated. Indeed, the shell structure cannot be determined uniquely from experimental data and cannot be talked about in an absolute sense as it depends on the non-observable resolution scale employed in the theoretical calculation. 
It is only at the price of \emph{fixing} arbitrarily (but conveniently!) such a scale that one can establish correlations between observables and the shell structure. To some extent, fixing the resolution scale provides ESPEs (and spectroscopic factors) with a \emph{quasi}-observable character. Eventually, practitioners can refer to nuclear shells and spectroscopic factors in their analyses of nuclear phenomena if, and only if, they use consistent structure and reaction theoretical schemes based on a fixed resolution scale they have agreed on prior to performing their analysis and comparisons. 
\end{description}
\end{abstract}


\date{\today}
\maketitle

\section{Introduction}
\label{s:intro}

The concept of single-nucleon shells dates back to the early days of contemporary nuclear physics and constitutes the basic pillar of the nuclear shell model~\cite{goeppertmayer49}. In fact, the notion of single-particle energy levels is a fundamental element of a large number of nuclear many-body theories and underlies our understanding of nuclear structure. Based on such a rationale, the correlated shell model has been able to explain the occurrence of extraordinarily stable configurations for specific neutron and proton numbers, known as magic numbers. The universal character of those magic numbers away from the valley of beta stability remains an open question~\cite{Sorlin08a} currently receiving considerable experimental and theoretical attention, e.g. see Refs.~\cite{baumann07a,janssens09a,Wienholtz:2013nya,Steppenbeck,Otsuka:2009cs,Holt:2010yb,Hagen:2012sh}. Whether a certain nucleon number qualifies as a (new) magic number cannot be postulated a priori. Experimentally, several quantities are measured to make such an assessment, e.g., the excitation energy and the collective character of the first $2^{+}$ state, the size of the gap in the one-nucleon addition/removal spectrum, etc.
Theoretically, the same quantities need to be computed by solving the many-body Schr\"odinger equation with sufficient accuracy in order to check whether the picture associated with a magic number holds. 

\begin{figure}
 \centering
  \includegraphics[width=.95\columnwidth,clip=]{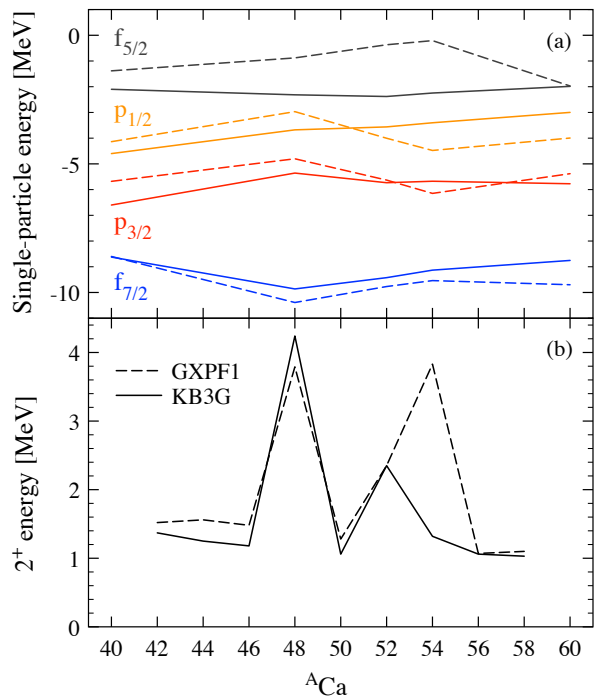}
\caption{(Color online). Shell-model calculation of selected Ca isotopes performed in the $pf$ valence space with GXPF1~\cite{honma04a} and KB3G~\cite{poves01a} empirical interactions. Upper panel: effective single-particle energies. Lower panel: first $2^{+}$ excitation energy.}\label{shellintro}
\end{figure}

As such, the characterization of the (non-)magic character of a nucleus is based on a set of many-body observables, and does \emph{not} involve the concept of single-nucleon shells. Still, effective single-particle energies~\cite{baranger70a} (ESPEs)\footnote{It must be made very clear that ``ESPE" refers throughout the present paper to the full Baranger-French definition of single-particle energies. In the traditional shell model, ``ESPE" usually refers to single-particle energies obtained by averaging over the monopole part of the Hamiltonian on the basis of a naive filling in an a priori given single-particle basis. The latter denotes an approximate version of the full Baranger-French definition obtained by omitting the correlations at play in the exact solution of the many-body problem.} associated with an auxiliary \emph{independent-particle-like} problem appear as a useful tool to interpret the evolution of many-body observables in terms of simpler theoretical ingredients. The typical picture is that the size of the first $2^{+}$ excitation energy of even-even isotopes as well as the size of the gap between their one-nucleon addition and removal spectra reflect (at least) qualitatively the size of the particle-hole gap in the ESPE spectrum and the value of the associated spectroscopic factors. For instance, a large gap at the Fermi energy in the ESPE spectrum is manifested as a large $2^{+}$ excitation energy with a reduced electric quadrupole transition probability $B(E2)$ to the ground state. Figure~\ref{shellintro} illustrates such a correlation for the first $2^{+}$ excitation energy of selected Ca isotopes on the basis of a $pf$ shell-model calculation. The phenomenological GXPF1~\cite{honma04a} and KB3G~\cite{poves01a} interactions both yield a high $2^{+}$ excitation energy and a large neutron $p_{3/2}-f_{7/2}$ shell gap in $\nuc{Ca}{48}$, but give very different predictions for the neutron-rich $\nuc{Ca}{54}$: GXPF1 predicts a large $f_{5/2}-p_{1/2}$ gap and high $2^{+}$ energy, KB3G the opposite. 

It is thus fair to say that the current paradigm underlining our understanding of nuclear structure provides the single-nucleon shell structure with a certain degree of ``reality''. As a matter of fact, reference to an underlying single-particle spectrum is almost systematically made to explain the characteristics and the evolution of low-energy observables in nuclei. Still, such a systematic reference raises basic questions given that individual nucleons do not occupy stationary single-particle states inside a correlated system. This relates to the fact that the only unambiguously defined problem that one can aim at addressing is the \emph{interacting} many-body problem, which translates into solving the A-body Schr\"odinger eigenvalue equation
\begin{equation}
H | \Psi^{\text{A}}_{k} \rangle =E^{\text{A}}_{k} | \Psi^{\text{A}}_{k}\rangle \label{Abodyschroe} \, \, ,
\end{equation}
and/or its time-dependent counterpart. While the outcome of the former takes the form of A-body energies $E^{\text{A}}_{k}$ and associated A-body states $| \Psi^{\text{A}}_{k}\rangle$, the latter provides reaction cross sections $\sigma(A_k\!+\!B_l\!\rightarrow\!C_m\!+\!D_n)$ associated with many-body systems transitioning from an initial state $| \Psi^{\text{A+B}}_{\text{initial}}\rangle = | \Psi^{\text{A}}_{k}\rangle  \otimes| \Psi^{\text{B}}_{l}\rangle$ to a final state $| \Psi^{\text{C+D}}_{\text{final}}\rangle = | \Psi^{\text{C}}_{m}\rangle\otimes| \Psi^{\text{D}}_{n}\rangle$. As such, single-nucleon shells do not appear explicitly in the formulation of the problem of interest. 

The closest accessible quantities relate to the dynamics of a nucleon that is added to or removed from the A-body correlated system, i.e., one-nucleon addition and removal energies $E^{\pm}_k \equiv \pm \big( E^{\text{A} \pm 1}_k - E^{\text{A}}_0 \big)$ along with associated\footnote{The shorthand notation $\sigma^{\pm}_k $ is used as a way to avoid specifying which reaction mechanism, e.g. which companion nucleus, is used to transfer a nucleon to/from the nucleus of interest.} reaction cross sections $\sigma^{\pm}_k $. As will be discussed in detail below, the computation of ESPEs, $e^{\text{cent}}_{p}(\lambda)$, combines one-nucleon addition and removal energies $E^{\pm}_k$ with associated spectroscopic probability matrices, $\mathbf{S}_{k}^{\pm}(\lambda)$. The dependence of ESPEs on the latter make them \emph{intrinsically} non-observable quantities that change with the resolution scale $\lambda$ employed in the theoretical description of the system. Thus, a unitary transformation of the Hamiltonian $H(\lambda)\rightarrow H(\lambda')$ changes the ESPE spectrum $e^{\text{cent}}_{p}(\lambda) \neq e^{\text{cent}}_{p}(\lambda')$ while leaving true observables invariant, e.g. $E^{\pm}_k(\lambda)=E^{\pm}_k(\lambda')$. Expanding on the brief discussion and the limited numerical illustrations provided in Ref.~\cite{Duguet:2011sq}, the goal of the present paper is to further characterize the non-observable character of the nuclear shell structure by presenting a detailed formal analysis along with systematic results obtained from ab initio many-body calculations.

The discussion proposed here is not meant to disqualify the notion of shell structure and the use of ESPEs in our interpretation of experimental data but to specify the terms, i.e., the exact sense and conditions, in which this can be done meaningfully. Still, it is crucial to state upfront that the non-observable character of the shell structure establishes that nuclear shells have no counterpart in the empirical world, i.e. in experiment, and that any apparent correlation between ESPEs and actual observables can only exist in a non-absolute sense. Indeed, ESPEs change with a ``parameter'' that is \emph{internal} to the theory and that can be tuned at will without modifying actual observables.  As will appear below, this ``parameter'' presently takes the form of a momentum scale $\lambda$ parameterizing families of unitary transformations that can be arbitrarily applied on the many-body Hilbert space.  While these unitary transformations do not modify the physics output, i.e., true observables\footnote{In the present text we use the same wording/notation to denote the self-adjoint operator associated with an observable and the corresponding eigenvalues accessed in a measurement.} $O$, they typically change any quantity that results from partitioning these observables, e.g. $O\equiv o_1(\lambda) + o_2(\lambda)$ or $O\equiv o_1(\lambda) \times o_2(\lambda)$. 

The non-observable nature of the one-nucleon momentum distribution~\cite{Furnstahl:2001xq}, of spectroscopic factors~\cite{Jennings:2011jm}, or of the one-nucleon shell structure~\cite{Duguet:2011sq} is not as esoteric or shocking as it may seem at first as it parallels situations encountered in other fields of physics. 
As a matter of fact, quantum mechanics and quantum field theories possess \emph{internal} degrees of freedom (e.g. the gauge symmetry) that are essential to their formulation but that are not observable, i.e., nothing in the empirical world can fix their value. Eventually, one can fix this freedom arbitrarily (and conveniently) such that non-observable quantities depending on it acquire a fixed value as well. Still, one must comply with the fact that the behavior of observables \emph{cannot} be correlated with non-observable quantities in an \emph{absolute} sense, but only when the internal degree of freedom is fixed to a particular value. Conversely, it is mandatory to agree on the way to fix this freedom prior to doing any comparison or even formulating any discourse on non-observable quantities~\cite{Furnstahl:2001xq,Duguet:2011sq}. The very same care associated with partitioning or factorizing observables is also routine in the discussion of parton distributions in hadronic physics (see, e.g., \cite{HandBookpQCD}).

\begin{figure}
 \centering
  \includegraphics[width=.95\columnwidth,clip=]{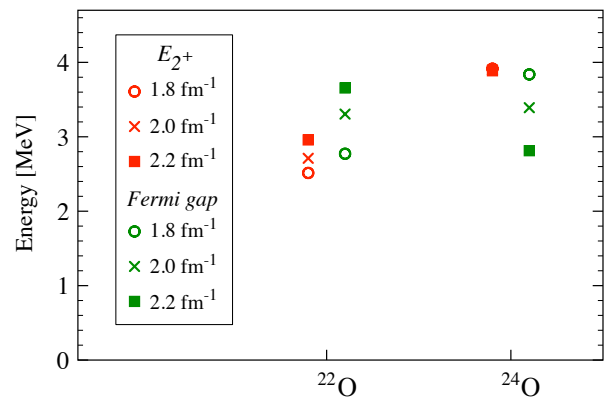}
\caption{(Color online). Comparison between the Fermi gap in the effective single-particle energy spectrum and the first $2^+$ excitation energy in $^{22,24}$O. 
Results are obtained from a microscopic shell model~\cite{Holt:2011fj,Caesar:2012tva} based on realistic 2N and 3N interactions. Calculations are displayed for three different values of the resolution scale characterizing the 2N interaction; see Ref.~\cite{Hebeler:2010xb,Simonis:2015vja} and the text for details.}\label{2plusFermigap}
\end{figure}

The non-observable character of ESPEs make them both resolution scale and theoretical scheme dependent. The present paper focuses on the former by studying at length the``running" of ESPEs with the scale $\lambda$ characterizing similarity renormalization group (SRG) transformations of the Hamiltonian~\cite{Bogner:2009bt}. Figure~\ref{2plusFermigap} anticipates this discussion by illustrating this feature from a microscopic shell model calculation ~\cite{Holt:2011fj,Caesar:2012tva} of $^{22,24}$O. This calculation is performed in a $sd$ valence space and is based on realistic two-nucleon (2N) and three-nucleon (3N) chiral effective field theory ($\chi$-EFT) interactions (see, e.g., \cite{Epelbaum:2008ga,Hammer:2013nx}) that are evolved to low momenta and further renormalized to the $sd$ shell through third-order many-body perturbation theory~\cite{Holt:2011fj}. The resolution scale characterizing the 2N interaction is varied from $1.8$ to $2.2$\,fm$^{-1}$ while keeping the regulator of the 3N interaction unchanged~\cite{Hebeler:2010xb,Simonis:2015vja}. 
In Fig.~\ref{2plusFermigap} the resulting values of the Fermi gap in the Baranger ESPE spectrum are compared with the first $2^{+}$ excitation energies. 
We observe that, while the ESPE Fermi gap typically changes by more than 1 MeV for both nuclei, the $2^+$ excitation energy varies by 400 keV in $^{22}$O and only 30 keV in $^{24}$O. Moreover, whereas the Fermi gap is relatively
close to the 2+ energy for the lowest resolution scale, it can
differ by up to 1 MeV for the highest one. Consequently, after varying the resolution scale the correlation 
between both quantities alluded to in connection with Fig.~\ref{shellintro} is 
weakened, while the observable $2^+$ excitation energy is unchanged.

The paper is organized as follows. Section~\ref{def} provides the detailed formal basis for the proper definition of ESPEs and to characterize their non-observable nature. Section~\ref{illustrations} illustrates the formal proofs via state-of-the-art many-body calculations based on $\chi$-EFT 2N and 3N interactions. Conclusions are given in Sec.~\ref{conclusions}.

\section{Nuclear shell energies}
\label{def}

\subsection{Rationale}

As already alluded to above, the interest of referring to single-nucleon shells resides in the hypothesis that low-energy observables reflect key patterns of the ESPE spectrum. Besides $2^{+}$ excitation energies, this is supposed to apply first and foremost to one-nucleon separation energies $E^{\pm}_k$. Such a rationale translates into the assumption that the observables can be partitioned into a dominant ``independent-particle-like'' component complemented by many-body correlations, i.e., that one can write schematically
\begin{equation}
\underset{\text{Outcome of Schr. equation}}{\underbrace{E^{\pm}_{k}}}  \, \, \, = \, \, \, \underset{\text{Ind. particles}}{\underbrace{e_{p}}} \, \, \, + \, \, \, \underset{\text{Correlations}}{\underbrace{\Delta E_{p\rightarrow k}}} \label{partitioning1} \, .
\end{equation}

Equation~\eqref{partitioning1} is a basic tenet of numerous many-body methods. For instance, in many-body perturbation theory, the independent-particle-like contribution refers to a chosen zeroth-order approximation. As such, it is the eigenvalue associated with an ad hoc one-body potential given a priori, e.g., of harmonic-oscillator (HO) or Woods-Saxon type. Alternatively, it can stem from a one-body potential that is derived from an auxiliary condition, 
e.g., the Hartree-Fock (HF) potential that results from minimizing the correlation contribution to the total binding energy. Another prime example is density functional theory (DFT), where single-particle energies are generated by a local one-body potential that emerges as a result of the constraint that the one-body local density of the Kohn-Sham Slater determinant matches the one of the exact A-body ground-state\footnote{Interestingly, such a constraint forces the single-particle energy of the last occupied Kohn-Sham orbital to match the one-fermion removal energy to the ground-state of the (A-1)-body system. This property is usually referred to as Koopmans'-like theorem of DFT. It, however, does not apply to any of the other Kohn-Sham single-particle energies that happen to have a non-trivial connection to one-nucleon separation energies~\cite{gross88a}. In practice, the validity of Koopmans'-like theorem of DFT is often compromised by spurious self-interaction problems~\cite{perdew81a}. Correcting for such an issue typically calls for \emph{orbital-dependent} density functionals~\cite{engel03a,Drut:2009ce}.}. Consequently, the single-particle energies typically discussed in the literature reflect a choice (among infinitely many) made by a practitioner. As such, they do not carry any deep meaning, and they certainly do not reflect a unique and unambiguous one-nucleon shell structure of the studied nucleus. Let us now introduce a superior definition of ESPEs~\cite{baranger70a,Duguet:2011sq}.

\subsection{Definition}
\label{barangerDEF}

The model-independent definition of effective single-particle energies relates them unambiguously to the process of adding (removing) a nucleon to (from) the ground-state of the A-body system of interest in (from) a specific single-particle state. The single-nucleon states in question are not known \emph{a priori}, but emerge together with ESPEs (see Eq.~\eqref{HFfield3}). 

We first specify the second-quantized form of the Hamiltonian entering Eq.~\eqref{Abodyschroe}. It is expressed in an arbitrary single-particle basis as
\begin{subequations}
\label{hamiltonian}
\begin{eqnarray}
H &=&  T + V^{\text{2N}} + V^{\text{3N}} + \ldots \label{hamiltonian1} \\
&=&\sum_{pq} t_{pq} a^{\dagger}_p a_q \nonumber \\
&&+ \left(\frac{1}{2!}\right)^2 \sum_{pqrs} \overline{v}^{\text{2N}}_{pqrs} a^{\dagger}_p a^{\dagger}_q a_s a_r \nonumber \\
&&+ \left(\frac{1}{3!}\right)^2 \sum_{pqrstu} \overline{v}^{\text{3N}}_{pqrstu} a^{\dagger}_p a^{\dagger}_q a^{\dagger}_r a_u  a_t a_s \nonumber \\
&&+ \ldots\, , \label{hamiltonian2}
\end{eqnarray}
\end{subequations}
where $\overline{v}^{\text{2N}}_{pqrs}$ and $\overline{v}^{\text{3N}}_{pqrstu}$ denote antisymmetrized matrix elements of 2N and 3N interactions while dots symbolize omitted higher-body forces. 

Next, we introduce the probability amplitudes $\mathbf{U}_{\mu}$ ($\mathbf{V}_{\nu}$) to reach a specific eigenstate $\ket {\Psi^{\text{A+1}}_{\mu}}$ ($\ket {\Psi^{\text{A-1}}_{\nu}}$) of the A+1 (A-1) system by adding (removing) a nucleon in (from) a single-particle state to (from) the ground state $\ket{\Psi^{\text{A}}_{0}}$ of an even-even system. Those amplitudes characterize \emph{direct} one-nucleon addition and removal processes and can be expanded in an arbitrary, e.g., spherical, single-particle basis $\{\ac {p}\}$ according to\footnote{Considering that $\ket{\Psi^{\text{A}}_{0}}$ is a $J^{\pi} = 0^+$ state and working with a spherical basis $\{\ac {p}\}$, i.e. $p\equiv(n,\pi,j,m,\tau)$, Wigner-Eckart's theorem states that the single-particle operator picks out the angular momentum, the parity and the isospin projection of the A$\pm$1 state the transfer goes to; i.e. $j_p=J_k$, $\pi=\Pi_k$ and $\tau=T_k-T_0$. Additionally, one can prove that $m=M$ ($-M$) for $U_k$ ($V_k$) where $M$ is the total angular-momentum projection of the A$\pm$1 state.}
\begin{subequations}
\label{eq:defu}
\begin{eqnarray}
U^{p}_{\mu} &\equiv& \bra {\Psi^{\text{A}}_{0}} a_p \ket {\Psi^{\text{A+1}}_{\mu}}  \, , \\
V^{p}_{\nu} &\equiv& \bra {\Psi^{\text{A}}_{0}} a^\dagger_p \ket {\Psi^{\text{A-1}}_{\nu}} \,  .
\end{eqnarray}
\end{subequations}
From these amplitudes, one builds spectroscopic probability matrices for the nucleon addition and removal, $\mathbf{S}_{\mu}^{+}\equiv \mathbf{U}_{\mu} \mathbf{U}^{\dagger}_{\mu}$ and $\mathbf{S}_{\nu}^{-}\equiv \mathbf{V}^{\ast}_{\nu}\mathbf{V}^{T}_{\nu}$, respectively. Their elements are
\begin{subequations}
\label{spectroproba}
\begin{eqnarray}
S_{\mu}^{+pq} &\equiv&  \bra {\Psi^{\text{A}}_{0}} a_p \ket {\Psi^{\text{A+1}}_{\mu}} \bra {\Psi^{\text{A+1}}_{\mu}} a^\dagger_q \ket {\Psi^{\text{A}}_{0}}    \, \, \, , \label{spectroprobaplus} \\
S_{\nu}^{-pq} &\equiv& \bra {\Psi^{\text{A}}_{0}} a^\dagger_q \ket {\Psi^{\text{A-1}}_{\nu}} \bra {\Psi^{\text{A-1}}_{\nu}} a_p \ket {\Psi^{\text{A}}_{0}}    \, \, \, . \label{spectroprobamoins}
\end{eqnarray}
\end{subequations}
Tracing the latter matrices over the one-body Hilbert space ${\cal H}_1$ provides \emph{spectroscopic factors}
\begin{subequations}
\label{spectrofactor}
\begin{eqnarray}
SF_{\mu}^{+} &\equiv& \text{Tr}_{{\cal H}_{1}}\!\left[ \mathbf{S}_{\mu}^{+}\right] =  \sum_{p \in {\cal H}_{1}} \left|U^{p}_{\mu}\right|^2 \, \, , \\
SF_{\nu}^{-} &\equiv& \text{Tr}_{{\cal H}_{1}}\!\left[\mathbf{S}_{\nu}^{-} \right] = \sum_ {p \in {\cal H}_{1}} \left|V^{p}_{\nu}\right|^2  \,\, ,
\end{eqnarray}
\end{subequations}
which are nothing but the norms of the spectroscopic amplitudes. A spectroscopic factor sums the probabilities that an eigenstate of the A+1 (A-1) system can be described as a nucleon added to (removed from) a single-particle state on top of the ground state of the A-nucleon system.

One can then gather the complete spectroscopic information associated with one-nucleon addition and removal processes into the so-called spectral function $\mathbf{S}(\omega)$. The spectral function denotes an energy-\emph{dependent} matrix defined on ${\cal H}_1$ through
\begin{eqnarray}
\mathbf{S}(\omega) &\equiv& \!\!\!\!\! \sum_{\mu \in {\cal H}_{A\!+\!1}} \!\!\! \mathbf{S}_{\mu}^{+} \,\, \delta(\omega -E_{\mu}^{+}) +  \!\!\!\!\!\sum_{\nu\in {\cal H}_{A\!-\!1}} \!\!\! \mathbf{S}_{\nu}^{-}  \,\, \delta(\omega -E_{\nu}^{-}) , \nonumber
\end{eqnarray}
where the first (second) sum is restricted to eigenstates of $H$ in the Hilbert space ${\cal H}_{A\!+\!1}$ (${\cal H}_{A\!-\!1}$) associated with the A+1 (A-1) system. Note that $\mathbf{S}(\omega)$ is directly related to the imaginary part of Dyson's one-body Green's function $\mathbf{G}(\omega)$~\cite{Dickhoff:2004xx}. Taking the trace of $\mathbf{S}(\omega)$ provides the spectral strength distribution (SDD)
\begin{eqnarray}
{\cal S}(\omega) &\equiv& \text{Tr}_{{\cal H}_{1}}\!\left[\mathbf{S}(\omega)\right]  \label{SDD_def} \\
&=& \! \!\!\!\!   \sum_{\mu\in {\cal H}_{A\!+\!1}} \!\!\!\!  SF_{\mu}^{+} \, \delta(\omega -E_{\mu}^{+}) + \sum_{\nu\in {\cal H}_{A\!-\!1}} \!\!\!\!    SF_{\nu}^{-}  \, \delta(\omega -E_{\nu}^{-}) \, ,\nonumber
\end{eqnarray}
which is a basis-independent function of the energy.

We also introduce the $n^{\text{th}}$ moment of the spectral function
\begin{equation}
\mathbf{M}^{(n)} \equiv \int_{-\infty}^{+\infty} \omega^{n} \, \mathbf{S}(\omega) \, d\omega , \label{spec_funct_moments}
\end{equation}
which defines an energy-independent matrix on ${\cal H}_1$. Using the anti-commutation rule of creation and annihilation operators $\{a_p,a^{\dagger}_q\} = \delta_{pq}$, the zero moment is shown to be nothing but the identity matrix
\begin{equation}
\mathbf{M}^{(0)}  = \sum_{\mu\in {\cal H}_{A\!+\!1}} \mathbf{S}_{\mu}^{+} + \sum_{\nu\in {\cal H}_{A\!-\!1}} \mathbf{S}_{\nu}^{-} = \mathbf{1} \, . \label{normalizationspectro}
\end{equation}
This sum rule provides each diagonal matrix element of $\mathbf{S}(\omega)$ with the meaning of a probability distribution function (PDF) in the statistical sense, i.e., the combined probability of adding a nucleon to or removing a nucleon from a specific single-particle basis state $| p \rangle$ integrates to 1 when summing over all the final states of the A$\pm$1 systems. 

\begin{figure}
 \centering
  \includegraphics[width=0.9\columnwidth,clip=]{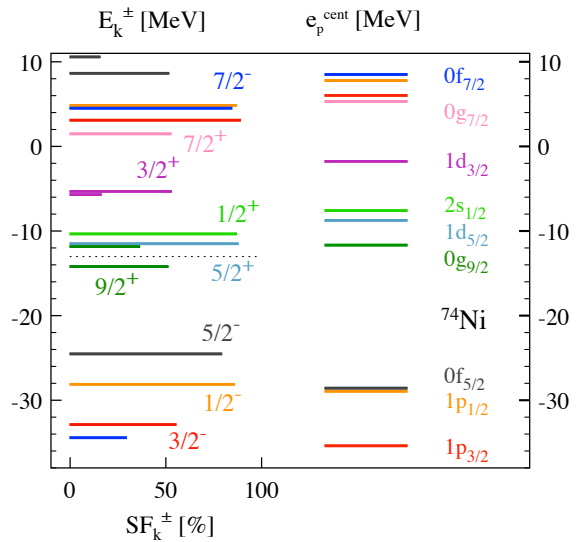}
\caption{(Color online). Self-consistent Gorkov Green's function calculation of $^{74}$Ni with a realistic 2N chiral interaction~\cite{entem03}. Left: spectral strength distribution for one-neutron addition (above the dashed line) and removal (below the dashed line) processes. Right: Baranger effective single-particle energies.}\label{spectra}
\end{figure}

The first moment $\mathbf{M}^{(1)}$ of the spectral function defines the so-called centroid matrix
\begin{eqnarray}
\mathbf{h}^{\text{cent}} &\equiv& \sum_{\mu\in {\cal H}_{A\!+\!1}} \mathbf{S}_{\mu}^{+} E_{\mu}^{+} + \sum_{\nu\in {\cal H}_{A\!-\!1}}  \mathbf{S}_{\nu}^{-} E_{\nu}^{-} \label{defsumrule} \,\,\, .
\end{eqnarray}
Effective single-particle energies are nothing but the \emph{eigenvalues} $\{e^{\text{cent}}_{p}\}$ of the centroid field~\cite{french66a,baranger70a}, and they are obtained by solving
\begin{eqnarray}
\mathbf{h}^{\text{cent}} \, \psi^{\text{cent}}_p &=& e^{\text{cent}}_{p} \, \psi^{\text{cent}}_p \,\,\, . \label{HFfield3}
\end{eqnarray}
Solving the eigenvalue problem~\eqref{HFfield3} not only provides ESPEs but also the corresponding single-particle states the nucleon is effectively added to or removed from. The associated spherical basis of ${\cal H}_1$ is denoted as $\{c^\dagger_{p}\}$. In that basis, ESPEs are expressed in terms of diagonal spectroscopic probabilities,
\begin{eqnarray}
e^{\text{cent}}_{p} &\equiv&  \sum_{\mu \in {\cal H}_{A\!+\!1}} S_{\mu}^{+pp} E_{\mu}^{+} + \sum_{\nu\in {\cal H}_{A\!-\!1}}  S_{\nu}^{-pp} E_{\nu}^{-}  \,\,\,. \label{HFfield2}
\end{eqnarray}
We see that ESPEs are nothing but centroids, i.e., an arithmetic average, of one-nucleon separation energies weighted by the probability to reach the corresponding A+1 (A-1) eigenstates by adding (removing) a nucleon to (from) a single-particle state $\psi^{\text{cent}}_p$. Centroid energies are by construction in one-to-one correspondence with states spanning ${\cal H}_1$. The step from one-neutron separation energies to neutron ESPEs is illustrated in Fig.~\ref{spectra} for an ab initio self-consistent Gorkov Green's function (G-SCGF) calculation~\cite{soma11a,Soma:2013xha} of $^{74}$Ni with a next-to-next-to-next-to-leading order (N$^3$LO) 2N chiral interaction~\cite{entem03} evolved down to a scale of $2$\,fm$^{-1}$ via a SRG transformation (see Sec.~\ref{illustrations} for details).

It is worth noting that Baranger ESPEs defined through Eqs.~\eqref{defsumrule}-\eqref{HFfield2} display three fundamental properties that make them fundamentally superior to {\it any} other definition of single-particle energies used in the literature: they (i) only invoke outputs of the many-body Schr\"odinger equation, (ii) do not depend on the single-particle basis used to expand the many-body problem and (iii) reduce to HF single-particle energies in the HF approximation, i.e., they satisfy Koopmans' theorem~\cite{koopmans34} in such a limit. Eventually, the model-independent character of Baranger ESPEs relates to the fact they can be computed unambiguously within any (re)formulation (i.e. scheme) of the nuclear many-body problem, e.g. shell model formulations, ab-initio formulations, cluster models etc. 

The fact that model-independent Baranger ESPEs {\it reduce} to HF single-particle energies in the HF {\it approximation} or to standard monopole ESPEs when employing a naive filling is best seen by applying the identity~\cite{vonbarth96a,vogt04a}
\begin{eqnarray}
M^{(n)}_{pq}  &=& \langle \Psi^{\text{A}}_0| \{\overset{n \, \text{commutators}}{\overbrace{[\ldots[[\ad{p},H],H],\ldots]}},\ac{q}\} |\Psi^{\text{A}}_0\rangle \, , \label{identitymoments}
\end{eqnarray}
to $n=1$~\cite{baranger70a,Polls94,Umeya:2006eh}
\begin{eqnarray}
h^{\text{cent}}_{pq} &=&  t_{pq} + \sum_{rs} \overline{v}^{\text{2N}}_{prqs} \, \rho^{[1]}_{sr} + \frac{1}{4}\sum_{rstv} \overline{v}^{\text{3N}}_{prtqsv} \, \rho^{[2]}_{svrt}  \label{HFfield} \\
&\equiv& h^{\infty}_{pq} \,\,\, , \nonumber
\end{eqnarray}
where
\begin{subequations}
\begin{eqnarray}
\rho^{[1]}_{pq} &\equiv&  \bra{\Psi^{\text{A}}_0} \ac{q} \ad{p} \ket{\Psi^{\text{A}}_0} = \sum_{\mu} {V^{p}_{\mu}}^{\ast} \, V^{q}_{\mu} \,\,\, , \\
\rho^{[2]}_{pqrs} &\equiv&  \bra{\Psi^{\text{A}}_0} \ac{r} \ac{s} \ad{q} \ad{p} \ket{\Psi^{\text{A}}_0}  \,\,\, ,
\end{eqnarray}
\end{subequations}
denote one- and two-body density matrices of the \emph{correlated} A-body ground-state, respectively. As Eq.~\eqref{HFfield} stipulates, the centroid field is equal to the one-body Hamiltonian $\mathbf{h}^{\infty} \equiv \mathbf{T}+\mathbf{\Sigma}(\infty)$ whose potential part is nothing but the energy-\emph{independent} component~\cite{Polls94} of the irreducible one-nucleon self-energy $\mathbf{\Sigma}(\omega)$ of the A-body ground state that naturally arises in self-consistent Green's-function theory. Equation~\eqref{HFfield} also makes clear that any many-body scheme capable of calculating $\rho^{[1]}$ and $\rho^{[2]}$ can extract the associated Baranger ESPEs, i.e. the definition is universal in that sense. Physically speaking, $\mathbf{h}^{\infty}$ represents the {\it average} one-body field seen by a nucleon {\it in presence} of correlations, i.e. not in a mean-field approximation. Taking such a simplified mean-field picture, e.g. the HF limit, one has
\begin{equation}
	\rho^{[2]}_{pqrs} = \rho^{[1]}_{pr} \rho^{[1]}_{qs} - \rho^{[1]}_{qr} \rho^{[1]}_{ps}\,,
\end{equation}
such that $\mathbf{h}^{\infty}$ reduces to the usual definition of $\mathbf{h}^{\text{HF}}$ for a 2N plus 3N Hamiltonian, 
which proves that $e^{\text{cent}}_{p} = e^{\text{HF}}_{p}$ in this limit.

\subsection{Partitioning of one-nucleon separation energies}

Let us now make the schematic partitioning introduced in Eq.~\eqref{partitioning1} more precise. We need to express one-nucleon separation energies in terms of ESPEs, i.e., invert Eq.~\eqref{HFfield2}. In order to achieve this goal, let us confront the eigenvalue equation providing ESPEs (Eq.~\eqref{HFfield3}) with the one satisfied by one-nucleon addition energies\footnote{A similar equation holds for one-nucleon removal energies $E^{-}_\nu$. Combining these equations of motion with Eq.~\eqref{normalizationspectro} to recover Eq.~\eqref{defsumrule} provides the identity $\sum_{\mu\in {\cal H}_{A\!+\!1}} \mathbf{\Sigma}^{\text{dyn}}(E^{+}_{\mu}) \, \mathbf{S}_{\mu}^{+} + \sum_{\nu\in {\cal H}_{A\!-\!1}} \mathbf{\Sigma}^{\text{dyn}}(E^{-}_\nu) \, \mathbf{S}_{\nu}^{-} = \mathbf{0}$.} (which derives from Dyson's equation~\cite{Dickhoff:2004xx})
\begin{eqnarray}
\left[\mathbf{h}^{\infty} + \left.\mathbf{\Sigma}^{\text{dyn}}(\omega)\right|_{\omega=E^{+}_{\mu}}\right] \mathbf{U}_{\mu} &=&  E^+_{\mu} \, \mathbf{U}_{\mu} \, , \label{confrontaa}
\end{eqnarray}
where $\mathbf{\Sigma}^{\text{dyn}}(\omega)\equiv \mathbf{\Sigma}(\omega)-\mathbf{\Sigma}(\infty)$ embodies the dynamical, i.e. energy-\emph{dependent}, part of the irreducible self-energy. Equation~\eqref{confrontaa} leads to
\begin{eqnarray}
E^+_{\mu} \, \text{Tr}_{{\cal H}_{1}}\!\left[ \mathbf{S}_{\mu}^{+}\right] &=&  \text{Tr}_{{\cal H}_{1}}\!\left[\mathbf{h}^{\text{cent}} \,\mathbf{S}_{\mu}^{+}\right] + \text{Tr}_{{\cal H}_{1}}\!\left[ \mathbf{\Sigma}^{\text{dyn}}(E^{+}_{\mu}) \,\mathbf{S}_{\mu}^{+} \right] \, , \nonumber
\end{eqnarray}
which, written in the centroid basis $\{c^\dagger_{p}\}$ diagonalizing $\mathbf{h}^{\infty}$, reads
\begin{eqnarray}
E^+_{\mu} &=& \sum_{p} s_{\mu}^{+pp} \, e^{\text{cent}}_p + \sum_{pq} s_{\mu}^{+pq} \, \Sigma^{\text{dyn}}_{qp}(E^{+}_{\mu})  \, . \label{partitioningexact2}
\end{eqnarray}
For each $\mu$, $\mathbf{s}_{\mu}^{+} \equiv \mathbf{S}_{\mu}^{+}/SF_{\mu}^{+}$ denotes the \emph{reduced} addition spectroscopic probability matrix whose trace over ${\cal H}_{1}$ is equal to one. For a given $\mu$, the set of diagonal matrix elements $\{\mathbf{s}_{\mu}^{+pp}\}$ thus possesses the meaning of a PDF. Equation~\eqref{partitioningexact2} provides a rigorous partitioning of one-nucleon addition energies into an independent-particle-like contribution and a correlation contribution.  Still, a given one-nucleon addition energy $E^{+}_{\mu}$ does not relate to a \emph{single} ESPE such that the connection between both spectra is actually of matrix character. The independent-particle-like contribution is the sum of all ESPEs weighted by the  probability for state $\ket {\Psi^{\text{A+1}}_{\mu}}$ to be obtained by adding a nucleon in the associated centroid states on top of $\ket {\Psi^{\text{A}}_{0}}$. The correlation contribution involves all matrix elements of the dynamical part of the  self-energy evaluated at the one-nucleon addition energy of interest. 

It is essential to stress that the partitioning \eqref{partitioningexact2} of the one-nucleon separation energy emerges as a direct consequence of the \emph{exact} Dyson equation \cite{Dickhoff:2004xx}. We have \emph{not made any assumption} regarding the way the many-body problem is solved to determine the ESPEs and the dynamical self-energy. At the same time, it is important to realize that the exact partitioning (\ref{partitioningexact2}) does not specify which effects are captured by the individual contributions, and we will demonstrate in the following that these details necessarily depend on the resolution scale $\lambda$.


\subsection{Non-observability}
\label{nonobserv}

We now come to the central point of the present study, i.e., the scale dependence and non-observable character of ESPEs. The latter derives directly from the scale dependence of spectroscopic amplitudes~\cite{Furnstahl:2001xq,Jennings:2011jm} entering the definition of the centroid Hamiltonian (Eq.~\eqref{defsumrule}).

\subsubsection{Description of low-energy nuclear systems}

So far, our discussion was conducted under the implicit assumption that one starts from a fixed and a priori given Hamiltonian $H$. In the context of low-energy nuclear physics, one is, from the outset, dealing with an effective theory of the underlying Quantum Chromo Dynamics (QCD). In this context, the paradigm is to employ $\chi$-EFT~\cite{weinberg91,ordonez94,vankolck94a} whose main merits are (i) to formulate the problem at hand in terms of relevant low-energy degrees of freedom (pions and nucleons) while retaining the (chiral) symmetry (breaking) of QCD, (ii) to provide a systematic framework for constructing all relevant interactions and the operators associated with other observables, and (iii) to explain the phenomenologically observed hierarchy of nuclear interactions, i.e., 2N interactions are more important than 3N interactions, which themselves dominate 4N forces, etc. This hierarchy is generated by the power counting that organizes the infinite set of interaction terms in the $\chi$-EFT Lagrangian~\cite{Bedaque:2002mn,Epelbaum:2008ga} according to their scaling with $(Q/\Lambda_{\chi})^{\nu}$. Here, $Q$ is a characteristic momentum scale for low-momentum processes and degrees of freedom, while $\Lambda_{\chi}$, the so-called chiral-symmetry-breaking scale, denotes the hard scale that characterizes omitted degrees of freedom and drives the low-energy constants in the chiral Lagrangian. $\chi$-EFT emphasizes that any self-adjoint operator associated with an observable, including the Hamiltonian, is \emph{effective} and thus depends on an intrinsic resolution scale $\Lambda_{\chi}$. At a given  order in $(Q/\Lambda_{\chi})^{\nu}$, operators of interest are schematically given by
\begin{equation}
O \equiv \sum_{\nu} O^{(\nu)}  \equiv O^{\text{1N}} + O^{\text{2N}} + \ldots + O^{\text{AN}}  \, \, \, , \label{defoperators}
\end{equation}
and necessarily contain up to A-body components as already alluded to above for the Hamiltonian. For instance, $V^{\text{2N}}$ first contributes to $H$ at leading order (LO) while $V^{\text{3N}}$ enters at next-to-next-to leading order (N$^{2}$LO) when using Weinberg's power counting (see, e.g., \cite{Bedaque:2002mn,Nogga:2005vn,Epelbaum:2008ga}).

Eventually, solving the A-body Schr\"odinger equation (Eq.~\eqref{Abodyschroe}) provides not only eigenfunctions $| \Psi^{\text{A}}_{k} \rangle$ and eigenenergies $E^{\text{A}}_{k}$ of $H$ but also other quantities of interest computed through, e.g., the average value $O^{\text{A}}_{k} \equiv \langle \Psi^{\text{A}}_{k}  | O | \Psi^{\text{A}}_{k}  \rangle$ of associated self-adjoint operators in a given eigenstate of $H$.

\subsubsection{Unitary transformation}

We consider a starting nuclear Hamiltonian $H$ built within $\chi$-EFT at a given order in the employed power counting. The Hamiltonian carries an intrinsic resolution scale characterized by both $\Lambda_{\chi}$ and the regularization cutoff(s) $\Lambda_{AN}$ that is introduced by the particular scheme used to renormalize many-body amplitudes at a given chiral order. 

Given this effective Hamiltonian, one is free to proceed to a unitary transformation $U(\lambda)$ over Fock space. The real variable $\lambda$ parametrizing the transformation typically denotes a momentum scale that characterizes the range of coupling between low and high momenta (within the interval defined by the intrinsic resolution scale of the starting $H$, which is itself a matter of choice) in the resulting Hamiltonian that takes the ``running'' form
\begin{subequations}
\begin{eqnarray}
H(\lambda) &\equiv& U(\lambda)H U^{\dagger}(\lambda) \equiv T + V^{\text{2N}}(\lambda) + V^{\text{3N}}(\lambda) + \ldots \, , \nonumber
\end{eqnarray}
\end{subequations}
where $V^{\text{AN}}(\lambda)$ changes with the scale $\lambda$. Even if the starting Hamiltonian $H$ were to contain only, e.g., one and two-body operators, its unitarily equivalent partner $H(\lambda)$ would in general contain (hopefully small) higher-body operators, which eventually truncate at the A-body level when applying the Hamiltonian on the A-body Hilbert space ${\cal H}_\text{A}$. Applying this unitary transformation to the Schr\"odinger equation, we obtain
\begin{eqnarray}
H(\lambda) | \Psi^{\text{A}}_{\mu} (\lambda) \rangle &=& E^{\text{A}}_k | \Psi^{\text{A}}_{\mu} (\lambda) \rangle \, , \label{transformedSchr}
\end{eqnarray}
where
\begin{eqnarray}
| \Psi^{\text{A}}_{\mu} (\lambda) \rangle &\equiv&  U(\lambda) \, | \Psi^{\text{A}}_{\mu} \rangle \, ,
\end{eqnarray}
such that the eigenvalues $E^{\text{A}}_k$ remain unchanged, while the many-body wave functions run with $\lambda$. Similarly, other operators transform under $U(\lambda)$ according to
\begin{eqnarray} 
O(\lambda) &\equiv& U(\lambda)\,  O \, U^{\dagger}(\lambda) \equiv O^{\text{1N}}(\lambda) + O^{\text{2N}}(\lambda) + O^{\text{3N}}(\lambda) + \ldots  \, . \nonumber
\end{eqnarray}
A key aspect of quantum mechanics concerns the assessment that the physical results, i.e. \emph{observables}, must remain unchanged under this unitary transformation. The consistent transformation of operators and many-body wave-functions ensures that eigenspectra of transformed operators, or more generally amplitudes of transformed operators between transformed states, including many-body cross sections, are indeed invariant under $U(\lambda)$~\cite{Schuster:2014lga}. 

An issue arises whenever a quantity is defined under the assumption that the associated operator should \emph{not} be transformed under $U(\lambda)$. This is the case for one-nucleon spectroscopic amplitudes that are \emph{defined} \emph{at any $\lambda$} as
\begin{subequations}
\label{eq:defu2}
\begin{align}
U^{p}_{\mu}(\lambda) &\equiv \bra {\Psi^{\text{A}}_{0}(\lambda)} a_p \ket {\Psi^{\text{A+1}}_{\mu}(\lambda)}  \, , \\
V^{p}_{\nu}(\lambda) &\equiv \bra {\Psi^{\text{A}}_{0}(\lambda)} a^\dagger_p \ket {\Psi^{\text{A-1}}_{\nu}(\lambda)}  \, ,
\end{align}
\end{subequations}
i.e., only the many-body states involved run with $\lambda$, not the operator. As a result, spectroscopic amplitudes undoubtedly vary with $\lambda$. One may suggest to transform the operator as well in the definition of spectroscopic amplitudes in order to make them invariant by construction. The transformed operator would have the general form
\begin{equation}
U(\lambda) \, a^\dagger_p \, U^{\dagger}(\lambda) = \sum_{q} u^{p}_{q}(\lambda) \, a^\dagger_q + \sum_{qrs} u^{p}_{qrs}(\lambda) \, a^\dagger_q a^\dagger_r a_s +\ldots \, , \label{transformedadditionop}
\end{equation}
with the initial conditions $u^{p}_{q}(\lambda_{\text{init}}) = \delta_{qp}$ for the first term and $u^{p}_{qrs\ldots}(\lambda_{\text{init}})=0$ for the others. Inserting such a form in the definition of the amplitudes would indeed lead to invariant spectroscopic factors and one-body centroid matrix $\mathbf{h}^{\text{cent}}$ (and thus ESPEs). However, the transformed operator~\eqref{transformedadditionop} clearly no longer corresponds to the addition of a nucleon in a specific single-particle state. Instead, it is a linear combination of not only one-particle operators but also two-particle/one-hole operators, three-particle/two-hole operators, etc. This contradicts the initial motivation behind the introduction of spectroscopic one-nucleon addition and removal amplitudes. Indeed, spectroscopic factors and ESPEs inform on the probability and the energy generated by adding and removing a nucleon through a process that involves a single nucleon state at a time, i.e., a pure direct process. If this is not the case, ESPEs no longer reduce to HF single-particle energies in the HF limit. To conclude, defining ESPEs in the context of a change of scale necessarily leads to keeping their definition formally the same for any $\lambda$ at the price of making their actual value scale dependent. The same goes for spectroscopic factors built from $\mathbf{U}_{\mu}(\lambda)$ and $\mathbf{V}_{\nu}(\lambda)$.

\subsubsection{Scale dependence}
\label{scaledep}

Following the spirit of the SRG~\cite{Bogner:2009bt}, a unitary transformation\footnote{It is common practice to test the predictions of different nuclear Hamiltonians for nuclei of interest. Ideally, all of these Hamiltonians are equivalent representations of low-energy QCD and describe observables like scattering data with high accuracy. In practice, however, traditional nuclear Hamiltonians have been derived using very different philosophies and theoretical frameworks. While their common link to QCD suggests implicit links between such Hamiltonians, there is no practical way to construct explicit transformations to study these connections. In contrast, the SRG provides a practical framework to build smoothly connected families of unitarily transformed nuclear Hamiltonians, and gives us a systematic handle on the violation of unitarity through truncations that are required in practical applications.} of the resolution scale can be defined through the differential flow equations of operators and many-body wave-functions
\begin{subequations}
\label{unitary2}
\begin{eqnarray}
\frac{d}{d\lambda} O(\lambda) &\equiv&  \left[ \eta(\lambda),O(\lambda)\right] \label{unitary2A} \, , \\
\frac{d}{d\lambda} | \Psi^{\text{A}}_{\mu} (\lambda) \rangle &\equiv&  \eta(\lambda) | \Psi^{\text{A}}_{\mu} (\lambda) \rangle \, , \label{unitary2B}
\end{eqnarray}
\end{subequations}
where the anti-hermitian generator of the transformation reads
\begin{equation}
\eta(\lambda) \equiv \frac{d U(\lambda)}{d\lambda} U^{\dagger}(\lambda) = -\eta^{\dagger}(\lambda) \, ,
\end{equation}
and the initial conditions are $O(\lambda_{\text{init}})=O$ and $| \Psi^{\text{A}}_{\mu} (\lambda_{\text{init}}) \rangle = | \Psi^{\text{A}}_{\mu} \rangle$. By combining Eqs.~\eqref{eq:defu2} and~\eqref{unitary2}, one obtains the flow equations for all quantities of interest. 

Starting from one-nucleon spectroscopic amplitudes
\begin{subequations}
\label{flowUandV}
\begin{eqnarray}
\frac{d}{d\lambda} V^{p}_{\nu}(\lambda) &=& - \langle \Psi^{\text{A-1}}_{\nu}(\lambda)| [\eta(\lambda),a_p] |\Psi^{\text{A}}_{0}(\lambda)\rangle^{\ast} \, , \label{flowUandVa} \\
\frac{d}{d\lambda} U^{p}_{\mu}(\lambda) &=& -\langle \Psi^{\text{A+1}}_{\mu}(\lambda)| [\eta(\lambda),a^\dagger_p] |\Psi^{\text{A}}_{0}(\lambda)\rangle^{\ast}  \, , \label{flowUandVb}
\end{eqnarray}
\end{subequations}
one obtains flow equations for spectroscopic probability matrices $\mathbf{S}_{\mu}^{+}$ and $\mathbf{S}_{\nu}^{-}$ as well as their traces $SF^{+}_{\mu}$ and $SF^{-}_{\nu}$. Combining Eq.~\eqref{flowUandV} with the fact that observable one-nucleon addition and removal energies are invariant because they are differences of eigenvalues of $H(\lambda)$ (see Eq.~\eqref{transformedSchr})
\begin{eqnarray}
\frac{d}{d\lambda} E^{-}_{\nu}(\lambda)  &=&  \frac{d}{d\lambda} E^{+}_{\mu}(\lambda)  =  0 \, , \label{flowE}
\end{eqnarray}
one can eventually derive flow equations for the zeroth and first moments of the spectral function matrix\footnote{Starting from Eq.~\eqref{identitymoments}, it is straightforward to derive the flow equation for an arbitrary moment $\mathbf{M}^{(n)}(\lambda)$.}
\begin{widetext}
\begin{subequations}
\label{flowMn}
\begin{eqnarray}
\frac{d}{d\lambda} \, M^{(0)}_{pq}(\lambda) &=&  0 \, , \label{flowMna} \\
\frac{d }{d\lambda} \, M^{(1)}_{pq}(\lambda) &=& - \langle \Psi^{\text{A}}_0 (\lambda) | \{[[\eta(\lambda),\ad{p}],H(\lambda)] ,\ac{q} \}+ \{[\ad{p},H(\lambda)] , [\eta(\lambda),\ac{q}] \} | \Psi^{\text{A}}_0 (\lambda) \rangle  \, . \label{flowMnb}
\end{eqnarray}
\end{subequations}
\vspace{0.7cm}
\end{widetext}
Equation~\eqref{flowMn} demonstrates that sum rule~\eqref{normalizationspectro} is scale invariant while the centroid matrix $\mathbf{h}^{\text{cent}}(\lambda)$ and its eigenvalues $e^{\text{cent}}_p(\lambda)$ are not.  Just as for spectroscopic factors, the latter property underlines the scale dependence of ESPEs, i.e., they ``run'' with the unitary transformation $U(\lambda)$, as opposed to true observables. 

\subsubsection{Symmetry transformations}

It is worth noting that symmetry transformations of the Hamiltonian associated with a (locally) compact Lie groups whose generators $C_i$ are one-body operators do \emph{not} induce any running of spectroscopic factors and ESPEs. Using a one-parameter group for simplicity and employing an exponential map to represent the transformation,  i.e., $U(\beta)=e^{i\beta C}$, one can show that the transformation of creation and annihilation operators in Fock space reduces to
\begin{subequations}
\label{symtransfo}
\begin{eqnarray}
U(\beta) \, a^{\dagger}_{p} \, U^{\dagger}(\beta) &=& \sum_{q} u_{qp}(\beta)  a_{q}^{\dagger} \, , \\
U(\beta) \, a_{p} \, U^{\dagger}(\beta) &=& \sum_{q} u^{\ast}_{qp}(\beta)  a_{q} \, ,
\end{eqnarray}
\end{subequations}
where $u_{qp}(\beta)\equiv \langle q | U(\beta) | p \rangle$ is the unitary matrix representing $U(\beta)$ in the one-body Hilbert space ${\cal H}_1$. In contrast to Eq.~\eqref{transformedadditionop}, a transformed creation (annihilation) operator remains a linear combination of pure one-particle creation (annihilation) operators. In this case, it is straightforward to show that spectroscopic probability matrices and the centroid matrix transform as standard matrices on ${\cal H}_1$ by using Eq.~\eqref{symtransfo}
\begin{subequations}
\begin{align}
S_{k}^{\pm pq}(\beta) &= \sum_{rs} u_{pr}(\beta) S_{k}^{\pm rs} u^{\dagger}_{sq}(\beta)  \, , \\
h^{\text{cent}}_{pq}(\beta) &= \sum_{rs} u_{pr}(\beta) h^{\text{cent}}_{rs} u^{\dagger}_{sq}(\beta) \, .\label{hcentsymmetry}
\end{align}
\end{subequations}
By virtue of the unitarity of $u_{qp}(\beta)$, the spectroscopic factors, i.e., the trace of the spectroscopic probability matrices, do not depend on $\beta$. Because $U(\beta)$ is a symmetry of $H$, it is also straightforward to show that $\mathbf{h}^{\text{cent}}$ is a \emph{scalar} and thus does not depend on $\beta$ either\footnote{Simply insert a complete basis of ${\cal H}_1$ whose states span the irreducible representations of the group in Eq.~\eqref{hcentsymmetry}.}.

Ultimately, this underlines the fact that we are presently not concerned with symmetry transformations. The transformations we are interested in are, e.g., free-space SRG transformations defined through their generator $\eta(\lambda)$ in such a way that the virtual coupling between low and high momenta is continuously reduced in $H(\lambda)$ (see Ref.~\cite{Bogner:2009bt} for details). Creation and annihilation operators are transformed on Fock space according to the general law~\eqref{transformedadditionop} and not the simpler transformation~\eqref{symtransfo}, which in turn causes spectroscopic factors and ESPEs to run with $\lambda$.
 
\subsubsection{Discussion}

The scale dependence of ESPEs generated by the flow equation~\eqref{flowMnb} has significant consequences. Despite the model-independent and physically intuitive character of Baranger's ESPEs, 
these quantities are not observable. Like spectroscopic factors, wave-functions or ``correlations'', nuclear shells do not qualify as an observable within the frame of quantum mechanics as they can be modified at will under a unitary transformation (while keeping true observables invariant). In that respect, the partitioning provided by Eq.~\eqref{partitioningexact2} can now be further specified as
\begin{widetext}
\begin{equation}
\overset{\text{many-body observable}}{\underset{\text{invariant under \,}  U(\lambda)}{\underbrace{E^+_{\mu}}}}    \equiv  \overset{\text{single-particle components}}{\underset{\text{varies under \,}  U(\lambda)}{\underbrace{\sum_{p} s_{\mu}^{+pp}(\lambda) \, e^{\text{cent}}_p(\lambda)}}}  \,\, +  \,\, \,\, \overset{\text{correlations}}{\underset{\text{varies under \,}  U(\lambda)}{\underbrace{\sum_{pq} s_{\mu}^{+pq}(\lambda) \, \Sigma^{\text{dyn}}_{qp}(E^{+}_{\mu};\lambda) \,}}} \label{partitioning2} \, ,
\end{equation}
\end{widetext}
which underlines that such a partitioning is necessarily scale dependent\footnote{The flow equation for the independent-particle-like contribution to one-nucleon separation energies can be easily worked out starting from Eqs.~\eqref{flowUandV} and~\eqref{flowMnb}. As the result is rather lengthy, we do not report it here.}.

An immediate consequence of the above analysis is the realization that extracting the single-particle shell structure and its evolution, e.g., with isospin, from experimental data is an illusory objective. The single-nucleon shell structure only exists \emph{within} the theoretical framework, given that experimental data determine the Hamiltonian \emph{only up to} a unitary transformation $U^{\dagger}(\lambda)U(\lambda)=1$. Any quantity that intrinsically depends on $\lambda$ is undefined in the empirical world and can only acquire the status of a \emph{quasi}-observable by fixing arbitrarily (but conveniently) the scale defining the Hamiltonian $H(\lambda)$ employed in the theoretical description. Still, any correlation established between an observable and a quasi-observable in an analysis performed at a particular scale may possibly disappear at another scale (see e.g. the example of Fig.~\ref{2plusFermigap}). For instance, the claim that a given one-nucleon addition energy $E^+_{\mu}$ tracks a specific particle-like ESPE $e^{\text{cent}}_p(\lambda)$ contributing to the first term on the right-hand side of Eq.~\eqref{partitioning2} can always be altered via a unitary transformation that leaves $E^+_{\mu}$ unchanged but that reshuffles the weight of the various terms on the right-hand side.
 
To fully appreciate the consequences of these statements, let us consider two practitioners independently analyzing the same (ideally) complete set of experimental data $\{E^{\pm}_{k} ; \sigma^{\pm}_{k}\}$ obtained from one-nucleon addition and removal experiments. Let us further postulate that they employ the same \emph{exact} structure and reaction many-body theories in their analyses but two different, though unitarily equivalent, Hamiltonians $H(\lambda)$ and $H(\lambda')$. As Eqs.~\eqref{flowE} and~\eqref{flowMn} demonstrate, the two practitioners will \emph{both} reproduce experimental observables perfectly, as $E^{\pm}_{k}(\lambda)=E^{\pm}_{k}(\lambda')$ and $\sigma^{\pm}_{k}(\lambda)=\sigma^{\pm}_{k}(\lambda')$, and fulfill sum-rule~\eqref{normalizationspectro}. However, they will extract two \emph{different} sets of nuclear shell energies and spectroscopic factors given that $e^{\text{cent}}_p(\lambda)\neq e^{\text{cent}}_p(\lambda')$ and $SF_{k}^{\pm}(\lambda)\neq SF_{k}^{\pm}(\lambda')$. This must be recognized prior to discussing any additional issue associated with approximations that necessarily come into play in practice. If not, the two practitioners are likely to conclude that the difference in the extracted shell energies and spectroscopic factors reflect those unavoidable approximations instead of realizing that in a perfect world they \emph{should} obtain \emph{different} results. We also note that the results can be compared directly (up to approximation uncertainties) if both practicioners agree on a scale and use transformation methods like the SRG to transform their results to this agreed-upon scale. This is a well-known application of RG methods, and common practice in the discussion of parton distribution functions in hadronic physics, e.g. see Ref.~\cite{HandBookpQCD}. 

\section{Illustrations}
\label{illustrations}

\subsection{Many-body calculations}
\label{manybody}

We perform many-body calculations for oxygen isotopes via state-of-the-art multi-reference in-medium SRG (MR-IM-SRG)~\cite{Tsukiyama:2010rj,Hergert:2012nb,Hergert:2013uja,Hergert:2014vn} and self-consistent Gorkov Green's function many-body methods~\cite{soma11a,Soma:2013xha}. From MR-IM-SRG we compute ground-state binding energies of even-even isotopes and  associated ESPEs. The same quantities are computed from G-SCGF along with one-nucleon addition and removal energies to eigenstates of the neighboring odd isotopes and their associated spectroscopic factors. Calculations employ a starting Hamiltonian containing a N$^3$LO \footnote{The denomination refers here to Weinberg's power counting.} 2N chiral interaction with a regularization cutoff $\Lambda_{\text{2N}}$=500 MeV~\cite{entem03} and a local N$^2$LO 3N interaction~\cite{Navratil2007tnf} with a  regularization cutoff $\Lambda_{\text{3N}}$=400~MeV. The 3N interaction low-energy constants $c_D$=-0.2 and $c_E$=0.098 are taken from a fit to the
ground-state energy and the $\beta$-decay half-life of $A=3$ 
systems~\cite{Roth2012prl}. Calculations are performed in model spaces of 14 (G-SCGF) and 15 (MR-IM-SRG) HO shells, respectively, which guarantees convergence with respect to the single-particle basis size to within 0.1\%. 
To manage the storage requirements of the 3N matrix elements, we only include configurations with $N_1$+$N_2$+$N_3\leq N^{\text{3N}}_{\text{max}}$=16 (G-SCGF) and 14 (MR-IM-SRG), where $N_i \equiv (2n_i+l+1)$ are HO single-particle energy quantum numbers. Uncertainties of G-SCGF and MR-IM-SRG due to $N^{\text{3N}}_{\text{max}}$ in the oxygen isotopes are significantly below 1\%, as discussed in Refs.~\cite{Soma:2014} and~\cite{Hergert:2012nb,Hergert:2013uja}, respectively.

The MR-IM-SRG calculations are performed in the so-called MR-IM-SRG(2) approximation that leads to well converged results from the many-body perspective~\cite{Hergert:2013uja}, at least as long as one employs interactions with low-momentum resolution scales. The G-SCGF calculations are performed at self-consistent second order~\cite{soma11a} in the Sc0 approximation~\cite{Soma:2013ona} that leads to well-converged results for relative quantities~\cite{Soma:2013xha} if low-momentum interactions are used. 

Our objective is to apply a sequence of SRG transformations $U(\lambda)$ to the starting chiral Hamiltonian and study the resulting variation of the quantities of interest. Ideally, we would like to vary $\lambda$ over the largest possible interval of values, i.e. from $\lambda_{\text{init}}$ to $0$. As illustrated in Sec.~\ref{breakingunitarity} below, we face two difficulties that result in breaking the unitarity of $U(\lambda)$, which complicates the illustration of the scale-dependence that is our main concern in the present work. On the one hand, many-body forces induced by the transformation need to be truncated beyond three-body operators when computing a A-body system with $A>3$. Technical limitations currently restrict free-space SRG transformation to relative two- and three-body Hilbert spaces \cite{Jurgenson:2009qs,Roth:2011ar,Hebeler:2012ly,Wendt:2013ys}, and the inclusion of four- and higher many-body forces in many-body calculations is an even bigger technical challenge. In order to prevent this truncation from inducing a too-significant breaking of unitarity, we have to keep the scale $\lambda$ of the Hamiltonian sufficiently large. On the other hand, the employed ab initio methods introduce truncations of the many-body expansion. At their current level of implementation, the resolution scale $\lambda$ of the Hamiltonian must be sufficiently small to ensure that the contribution of the truncated terms of the many-body expansion does not exceed the desired accuracy for the calculated observables, which is on the level of 2-3\% in oxygen. To accommodate these two opposing constraints, we are currently forced to work in a limited interval of resolution scales encompassing the three values $\lambda = 1.88, 2.00, 2.24$\,fm$^{-1}$ used in this work. Nevertheless, we will demonstrate below that this range is sufficient to illustrate the formal arguments provided above. Obviously, the quality of the illustrations proposed below can be improved in the future if (i) the SRG transformation can be conducted in Hilbert spaces ${\cal H}_A$ with $A>3$, (ii) four- and higher-body forces can be handled in many-body calculations and if (ii) the many-body truncations enforced within MR-IM-SRG and G-SCGF theories reach a maturity ensuring the convergence of calculations based on Hamiltonians characterized by a large resolution scale $\lambda$.
 
 \begin{figure}[t]
 \centering
  \includegraphics[width=0.99\columnwidth,clip=]{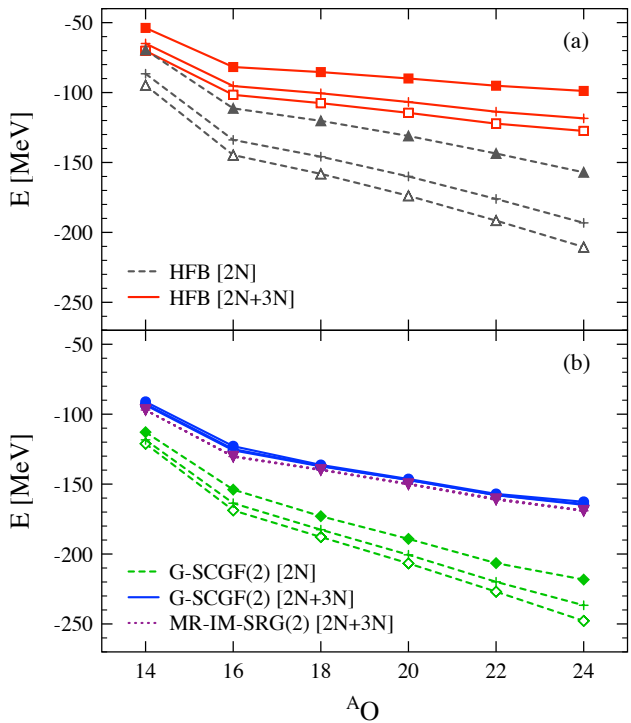}  
\caption{(Color online). Ground-state binding energies of $^{14-24}$O computed for $\lambda = 1.88$ (open symbols), $2.00$ (crosses), and $2.24$ fm$^{-1}$ (filled symbols). Panel (a): lowest order (i.e. HFB) calculation for two cases: keeping the sole 2N interaction and incorporating both starting and induced 3N interactions. Panel (b): G-SCGF second-order results without and with 3N forces (as above) and MR-IM-SRG(2) calculation with the full 2N+3N Hamiltonian.}
\label{breaking1}
\end{figure}

\subsection{Breaking unitarity}
\label{breakingunitarity}

In order to qualify the intrinsic running of non-observable ESPEs and spectroscopic factors, one must quantify the (presently) unavoidable, though artificial, scale dependence of observables in any practical calculation. As already alluded to, the latter originates from the breaking of unitarity due to (i) the omission of induced many-body forces beyond $A=3$ and (ii) truncations of the many-body expansion when the many-body Schr\"odinger equation is solved.

Given that we use a finite interval of scale variation, it is difficult to characterize the extent of this artificial scale dependence of observables in an absolute sense. In order to provide some perspective, we first alter our G-SCGF  calculations by retaining 2N interaction only\footnote{Providing the results obtained while only retaining 2N interaction also allows to relate to the calculations that were possible at the time where Ref.~\cite{Duguet:2011sq} was published and illustrate the benefit brought about by the inclusion of three-body forces in coupled-cluster, self-consistent Green's function and in-medium similarity renormalization group that was achieved in the meantime~\cite{Hagen:2007ew,Binder:2013oea,Hergert:2013uja,Carbone:2013eqa,Cipollone:2013zma,Soma:2013vca,Soma:2013xha,Binder:2013xaa}.} (thus omitting all many-body interactions induced from it including the 3N operator) and/or by degrading the self-consistent second-order treatment to the strict Hartree-Fock-Bogoliubov (HFB) level that acts as a zeroth-order baseline. The corresponding results are displayed in Fig.~\ref{breaking1} for the ground-state binding energy of $^{14-24}$O.

\begin{figure*}
 \centering
  \includegraphics[width=2.0\columnwidth,clip=]{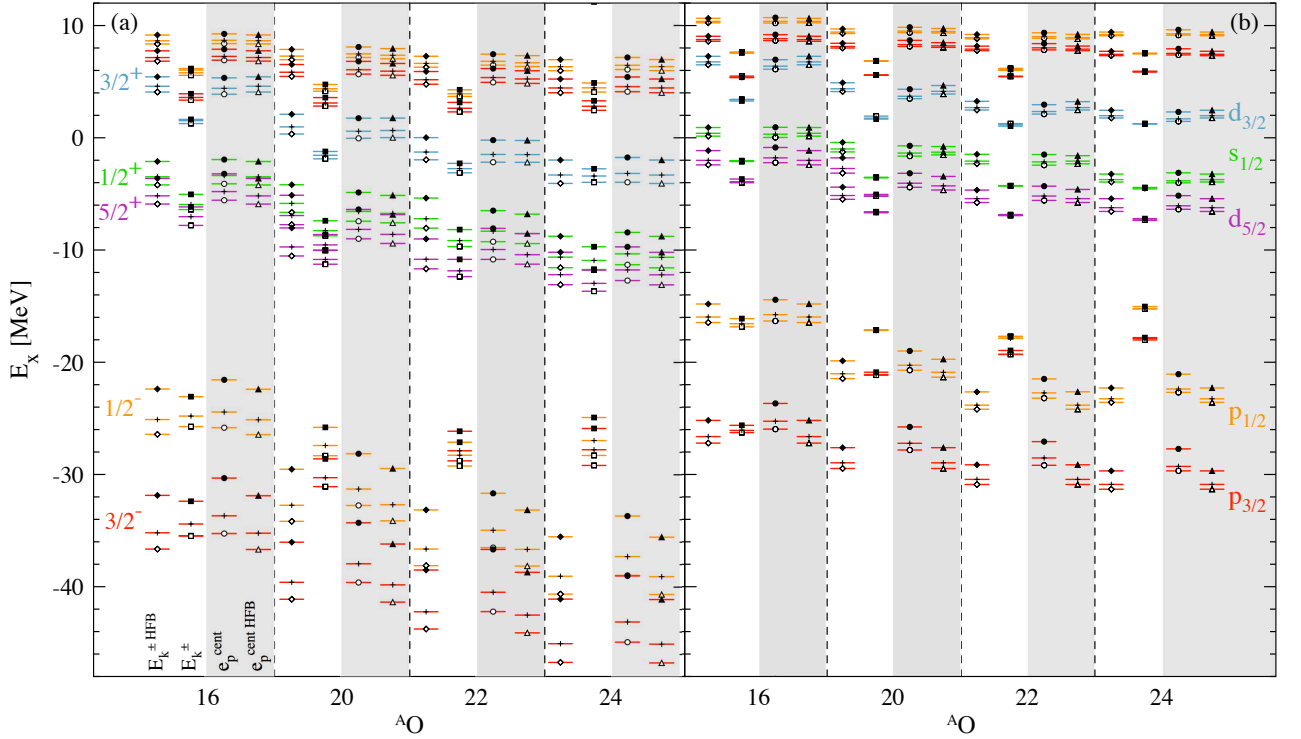}
\caption{(Color online) One-neutron separation energies with dominant spectroscopic factors versus neutron ESPEs in $^{16,20,22,24}$O. Each level is displayed for $\lambda = 1.88$ (open symbols), $2.00$ (crosses), and $2.24$\,fm$^{-1}$ (filled symbols). Results are displayed for both HFB and second-order G-SCGF calculations. Panel (a): one- and two-body operators are retained in the initial and transformed Hamiltonians. Panel (b):  one-, two-, and three-body operators are retained in the initial and transformed Hamiltonians.}
\label{running1}
\end{figure*}

First, the scale dependence generated over the interval $\lambda \in [1.88,2.24]$ is systematically smaller in the calculations that keep 2N and 3N operators than in the calculation only retaining the 2N interaction\footnote{The most meaningful comparisons (not given here) would consist in having the calculation keeping 2N and 3N operators while omitting the original 3N interaction in the chiral Hamiltonian~\cite{Binder:2013oea,Hergert:2013uja,Carbone:2013eqa,Cipollone:2013zma,Soma:2013vca,Soma:2013xha,Binder:2013xaa}. In the present case we only interpret as indicative the reduction of the running with $\lambda$ because the calculation contains the original 3N interaction in addition to the induced one.}. Second, the reduced scale dependence is very significant in both calculations when going from HFB to self-consistent second order. Using 2N+3N forces for example, the scale dependence is reduced from about 20-30\,MeV to typically 2\,MeV when spanning the (small) interval $\lambda \in [1.88,2.24]$\,fm$^{-1}$, i.e. by a factor $\sim15$. Note, however, that part of the reduction is due to a cancellation between induced 4N interactions from the initial 2N and 3N interactions, as discussed in Refs.~\cite{Binder:2012mk,Hergert:2012nb,Hergert:2013uja,Binder:2013xaa}.

In order to verify that the pattern just discussed is not specific to G-SCGF but reflects a generic aspect of the many-body problem, we further compare in panel (b) of Fig.~\ref{breaking1} with MR-IM-SRG(2) calculations for the Hamiltonian containing 2N+3N forces. At the current level of implementation, the MR-IM-SRG includes many-body terms beyond G-SCGF, and allows an even more significant reduction of the scale dependence, while also benefitting from the cancellation of induced 4N terms mentioned above. The residual running ranges from $50$\,keV in ${}^{14}\mathrm{O}$ to $400$\,keV in ${}^{24}\mathrm{O}$ for $\lambda \in [1.88,2.24]$\,fm$^{-1}$. 
The better many-body convergence of MR-IM-SRG(2) is also reflected in the additional absolute binding~\cite{Soma:2013xha, Hergert:2014vn}. 
A third-order G-SCGF truncation scheme will provide the missing binding energy and will allow for a further attenuation of the scale dependence, as shown in Ref.~\cite{Cipollone:2013zma} for closed-shell oxygen isotopes.

\subsection{Nuclear shell energies}
\label{running}

\begin{figure}[t]
 \centering 
 \includegraphics[width=0.99\columnwidth,clip=]{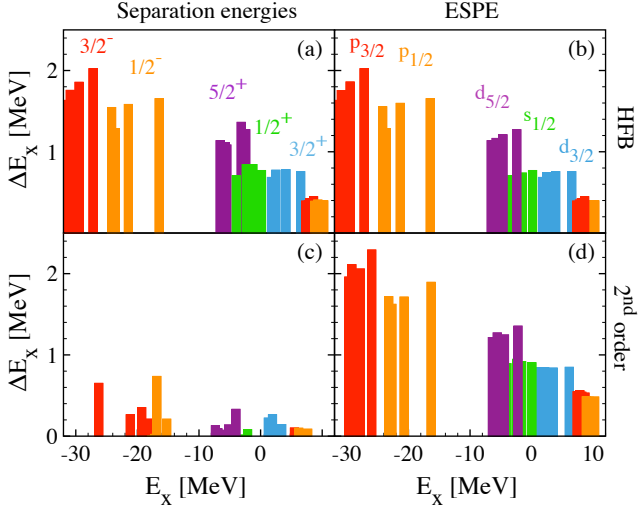}
\caption{(Color online) Residual spreads of separation energies and ESPEs in HFB and second-order G-SCGF calculations with a 2N+3N Hamiltonian. Differences between $\lambda = 1.88$ and $2.24$\,fm$^{-1}$ calculations are displayed for each spin-parity state. For one-neutron separation energies, states with spectroscopic factor larger than $30\%$ are retained. All states displayed in Fig.~\ref{running1}(b) are plotted indistinctly. Going from panel (a) to panel (c), one notices (i) a large reduction of the scale dependence and (ii) the expected compression of the strength due to the inclusion of the coupling to fluctuations. Comparing panels (b) and (d) makes clear that none of these two features is reflected in the ESPEs, i.e. in the underlying shell structure.}
\label{spreads}
\end{figure}

First, we compare one-nucleon separation energies $E^{\pm}_k$ with absolute ESPEs $e^{\text{cent}}_p$ in $^{16,20,22,24}$O. For each spin and parity, we consider the separation energy of the state with the dominant strength\footnote{The two visible $5/2^+$ levels in $^{20}$O actually correspond to two different states with similar strength. The fact that two states with equal strength appear near the Fermi energy is characteristic of the superfluid and open-shell nature of $^{20}$O.}. As in the previous section, we perform HFB and G-SCGF calculations using the SRG-evolved 2N and 2N+3N Hamiltonians, and compile results from all four variants in Fig.~\ref{running1}, covering energies from $-48$\,MeV to $+10$\,MeV. Let us now list the main lessons one can learn from these results.
\begin{itemize}
\item Combining panels (a) and (b), one can appreciate the significant reduction of the artificial scale dependence of \emph{all} one-nucleon separation energies obtained by keeping 3N operators in the Hamiltonian and/or by going from HFB to second-order G-SCGF.
\item The running of ESPEs is qualitatively different and quantitatively larger than for observable one-nucleon separation energies. This is particularly clear for the 2N+3N Hamiltonian: While the average spread of all displayed separation energies is equal to $0.2$\,MeV for $\lambda \in [1.88,2.24]$\,fm$^{-1}$, the average spread of ESPEs is equal to $1.1$\,MeV. The distribution of those spreads is shown in detail in Fig.~\ref{spreads}.
\item While the spread of ESPEs is qualitatively different and quantitatively larger than that of observable one-nucleon separation energies, it is worth mentioning that the latter are much more sensitive to correlations than the former. Indeed, while the values of a given separation energy computed for $\lambda \in [1.88,2.24]$\,fm$^{-1}$ converge towards one another when going from HFB to self-consistent second-order, each of them does so by changing significantly on an absolute scale. This systematic convergence of the one-nucleon separation energies computed for three different scales to a common value as one improves the many-body treatment is not fortuitous but rather reflects the intrinsic scale independence of these observables. This trend is qualitatively different for ESPEs whose genuine spread ultimately remains similar (and significant) when going from HFB to self-consistent second order.
\end{itemize}

\begin{figure}
 \centering
  \includegraphics[width=0.99\columnwidth,clip=]{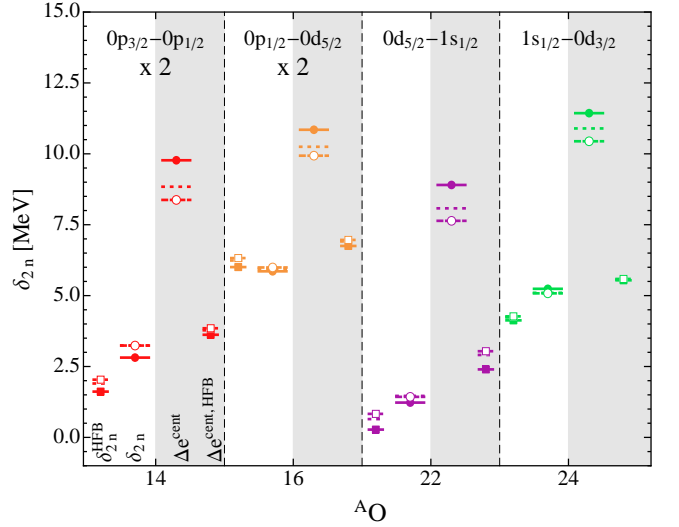}
\caption{(Color online) Two-nucleon shell gap versus the ESPE Fermi gap in $^{14,16,22,24}$O. Each quantity is displayed for $\lambda = 1.88$ (dashed lines, open symbols), 2.00 (dotted lines, no symbols), and 2.24 \,fm$^{-1}$ (solid lines, filled symbols).  Results are displayed for both HFB ($\delta^\text{HFB}_{2n}, \Delta e^\text{cent,HFB}$) and the MR-IM-SRG(2) ($\delta_{2n}, \Delta e^\text{cent}$) truncation scheme. One-, two-, and three-body operators are retained in the initial and transformed Hamiltonians. }
\label{running2}
\end{figure}

Second, we compare in Fig.~\ref{running2} HFB and MR-IM-SRG(2) results for the so-called two-neutron shell gap and the ESPE gap across the Fermi energy, focusing on the nuclei $^{14,16,22,24}$O that display good closed sub-shell character for the 2N+3N Hamiltonian (e.g., there is no pairing at the HFB level). The observable two-neutron shell gap and the ESPE gap are defined through \cite{Bender:2005ri,Bender:2008zr}
\begin{align}
\delta_{2n}(N,Z)&\equiv \frac{1}{2}\left( E(N\!+\!2,Z) - 2 E(N,Z) + E(N\!-\!2,Z)\right)\, ,
\end{align}
and
\begin{align}
\Delta e^\text{cent}(N,Z)&\equiv e^\text{cent}_p(N,Z) - e^\text{cent}_h(N,Z)\,,
\end{align}
respectively. By definition $e^\text{cent}_h(N,Z)$ ($e^\text{cent}_p(N,Z)$) denotes the ESPE energy of the last occupied (first empty) shell obtained via a naive, i.e. non interacting, filling of those shells for the even-even system with $N$ neutrons and $Z$ protons.
Disregarding the change in the single-particle wave functions when going from $N$ to $N\pm 2$ nuclei, along with the interaction between the added (removed) two neutrons, it is easy to see that $\delta_{2n}(N,Z)$ and $\Delta e^\text{cent}(N,Z)$ are equal in the HF limit. This is the reason why the former observable is often compared to the latter non observable ESPE gap.

The main lessons to retain from Fig.~\ref{running2} are similar to before.
\begin{itemize}
\item As expected from good doubly-closed shell systems, the ESPE Fermi gap captures the two-neutron shell gap quantitatively at the mean-field, i.e. HFB, level independently of the scale used. Contrarily, this is not at all the case at the MR-IM-SRG(2), i.e. correlated, level. This is typical of ab-initio theoretical schemes where the dynamics of all nucleons is treated on the same footing, as was already exemplified above for one-nucleon separation energies from second-order G-SCGF calculations as well as from CC calculations at the singles and doubles level in Ref.~\cite{Duguet:2011sq}. 
\item The scale dependence of the ESPE Fermi gap is qualitatively different and systematically larger than the artificial running of the two-neutron shell gap, thus illustrating the non-observable (observable) nature of the former (latter). The scale dependence of $\delta_{2n}(N,Z)$ is reduced systematically from $600-700\,\keV$ to $200\,\keV$ by going from HFB to MR-IM-SRG(2), with the exception of $\nuc{O}{14}$, where the $\delta_{2n}(N,Z)$ obtained from HFB and MR-IM-SRG(2) are both $\sim850\,\keV$. In contrast, the scale dependence of the ESPE Fermi gap grows from $400-600\,\keV$ to $1.5-2.8\,\MeV$ in $\nuc{O}{14,16,22}$ as we go from HFB to MR-IM-SRG(2)\footnote{Superficially, the increase in the scale dependence is even more severe for $\nuc{O}{24}$, where the gap increases from $40\,\keV$ to $1\,\MeV$. However, continuum effects, which are currently only taken into account through the crude discretization provided by the HO basis, should be especially important in that nucleus.}. 
\end{itemize}

The above results constitute the best illustration currently allowed by state-of-the-art many-body calculations of the scale dependence of nuclear shell energies. While we consider this illustration to be already striking, its quality will keep improving over the coming years, as already mentioned in Sect.~III.B.

\subsection{Spectroscopic factors}
\label{SFactors}

\begin{figure}
 \centering
  \includegraphics[width=0.99\columnwidth,clip=]{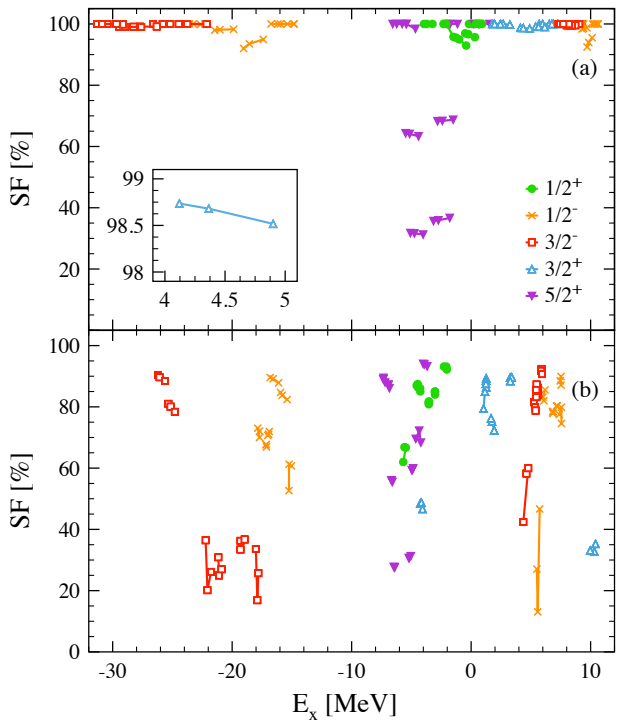} \\
\caption{(Color online) Spectroscopic factors associated with one-neutron addition and removal process on the ground states of  $^{14,16,18,20,22,24}$O computed as a function of the associated separation energy. For each final state, results obtained for $\lambda = 1.88, 2.00, 2.24$\,fm$^{-1}$ are joined by solid lines. One-, two-, and three-body operators are retained in the (initial and) transformed Hamiltonians. Panel (a): results obtained at the HFB level. Panel (b): results obtained from second-order G-SCGF calculations.}
\label{running3}
\end{figure}

Let us also briefly illustrate the non-observable character of spectroscopic factors. To do so, spectroscopic factors associated with one-neutron addition and removal processes on the ground states of  $^{14,16,18,20,22,24}$O are compiled in Fig.~\ref{running3} as a function of the separation energy of the corresponding final state. For each state, the results obtained for $\lambda = 1.88, 2.00, 2.24$\,fm$^{-1}$ are connected by lines. At the HFB level (Fig.~\ref{running3}(a)), the variation of the spectroscopic factors with $\lambda$ is sufficiently small to be obscured by the symbols. This variation essentially occurs horizontally because the one-neutron separation energies do depend on $\lambda$ (see inset in Fig.~\ref{running3}(a)) at that level as discussed previously. Contrarily, there is essentially no vertical variation as one is operating within an independent (quasi)-particle picture such that the strength for particle addition or removal is contained almost entirely in individual single-particle states, i.e. the eigenstates of $\mathbf{h}^\text{cent}$, by construction. Spectroscopic factors are actually not strictly equal to 1 (or 0) due to the treatment of pairing correlations in the HFB framework, the two $5/2^+$ states associated with one-neutron addition and removal in the open-shell $^{20}$O being the most prominent example. 

The picture is different in Fig.~\ref{running3}(b), where results from second-order G-SCGF calculations are compiled. There is less variation along the horizontal axis than in the HFB case because the improved many-body treatment reduces the scale dependence of the \emph{observable} one-neutron separation energies (cf.~Fig.~\ref{running1}). Due to the inclusion of dynamical correlations, the spectroscopic strength is now fragmented. For certain states, in particular the $1/2^-$ and $3/2^-$ states, the vertical spread becomes visible and indicates that the details of this fragmentation depend on the resolution scale $\lambda$ (while the associated separation energy does not). By improving the treatment of the many-body problem through switching from HFB to G-SCGF, we have thus slightly increased the scale dependence of some of the \emph{non-observable} spectroscopic factors significantly. Still, a larger range of $\lambda$ values will have to be used in order to generate any significant and systematic scale dependence. 

\section{Conclusions}
\label{conclusions}

The present work is dedicated to specifying and illustrating the non-observable nature of the one-nucleon shell structure. After a formal demonstration, state-of-the-art multi-reference in-medium similarity renormalization group and self-consistent Gorkov Green's function many-body calculations based on chiral two- and three-nucleon interactions are employed to illustrate that, as opposed to observable quantities, nuclear shell energies run under unitary similarity renormalization group transformations of the Hamiltonian parameterized by the resolution scale~$\lambda$. 
In practice, the unitarity of the similarity transformations is broken due to the omission of induced many-body interactions in the present framework, and the approximate treatment of the Schr\"odinger equation. The impact of this breaking is first characterized by quantifying the (artificial) running of observables over a (necessarily) finite interval of $\lambda$ values. Then, the (genuine) running of ESPEs is characterized and shown to be convincingly larger than that of observables (which would be zero in an exact calculation). 

The non-observable nature of the nuclear shell structure, i.e., the fact that it constitutes an intrinsically theoretical object with no counterpart in the empirical world, must be recognized and assimilated. Indeed, the shell structure cannot be extracted from experimental data; hence it cannot be talked about in an absolute sense as it depends on the non-observable resolution scale employed in the theoretical calculation. Consequently, correlations that one may establish between observables, e.g., first $2^{+}$ excitation energies or one-nucleon separation energies, and features of the shell structure, e.g., the size of the particle-hole gap at the Fermi energy, depend on the resolution scale. It is only at the price of \emph{fixing} arbitrarily (but conveniently!) the resolution scale in the theoretical framework that one can establish and utilize such correlations. To some extent, fixing the resolution scale provides ESPEs (and spectroscopic factors) with a \emph{quasi}-observable character. 

Ultimately, practitioners can refer to nuclear shells and spectroscopic factors in their analyses of nuclear phenomena. This however requires that it is done on the basis of a well defined theoretical scheme, i.e. well specified degrees of freedom combined with a Hamiltonian characterized by a fixed resolution scale. It is mandatory to perform comparisons from one nucleus to the other or from one practitioner to the other on the basis of that {\it very same theoretical scheme}. Incidentally, this also necessitates to use \emph{consistent} structure and reaction theoretical schemes, i.e. structure and reaction theories based on the same degrees of freedom and the same fixed Hamiltonian, eventually employing the same approximations within that many-body scheme. This is of course a very challenging task for the future. Still, it indicates that, from the perspective of future theoretical developments, there is not much value in combining, e.g., high-quality ab initio nuclear structure quantities with inconsistent nuclear reaction theories. The focus should rather be on \emph{consistency} as there is more value in developing less advanced, e.g. less ab initio, structure and reaction theories \emph{as long as} the degrees of freedom, the many-body truncation schemes and the nuclear Hamiltonian underlying both are consistent.

\begin{acknowledgments}
This work was initiated during a programme of Espace de Structure Nucl\'eaire Th\'eorique (ESNT) at CEA Saclay, from which the authors acknowledge support.
%
T.~D. wishes to thank G.~Hagen for his early collaboration on the topic of present interest. We thank C.~Barbieri and R.~J.~Furnstahl for useful discussions. We thank J.~Men\'endez, K. Hebeler, A.~Schwenk, and J.~Simonis for their collaboration on the NN+3N shell-model calculations, C. Barbieri and A. Cipollone for their collaboration on the G-SCGF calculations, S.~Bogner and T.~Morris for their collaboration on the MR-IM-SRG, as well as S.~Binder, A.~Calci, J.~Langhammer, P.~Navr\'atil, and R.~Roth for providing us with matrix elements of SRG-evolved chiral 3N interactions.
H.~H.~acknowledges support by the National Science Foundation (NSF) under Grant No.~PHY-1306250, and the NUCLEI SciDAC-3 Collaboration under the U.S. Department of Energy Grant No.~DE-SC0008533. 
J.~D.~H. was supported by the BMBF Contract No.~06DA70471, the Helmholtz Alliance HA216/EMMI, the National Research Council of Canada and NSERC.
G-SCGF calculations were performed using HPC resources from GENCI-CCRT (Grant no. 2014-050707). 
Computing resources for MR-IM-SRG calculations were provided by the Ohio Supercomputing Center (OSC), and the Michigan State University High Performance Computing Center (HPCC)/Institute for Cyber-Enabled Research (iCER). Many-body perturbation theory computations were performed on JUROPA at the J\"ulich Supercomputing Center.

\end{acknowledgments}

\bibliography{shell_structure}

\end{document}